\documentclass[11pt]{article}
\usepackage{epsfig}
\usepackage{amssymb,amsmath,amsthm,amscd}
\usepackage{latexsym}

\newcounter{labelflag} 

\newcommand{\Label}[1]{
                       \ifnum\thelabelflag=1
                          \ifmmode
                             \makebox[0in][l]{\qquad\fbox{\rm#1}}
                          \else
                             \marginpar{\vspace{0.7\baselineskip}
                                        \hspace{-1.1\textwidth}
                                        \fbox{\rm#1}}
                          \fi
                       \fi
                       \label{#1}
                      }
\newcommand{\be}{\begin{equation}}
\newcommand{\ee}{\end{equation}}

\newcommand{\nb}{ Newton-Boussinesq equation }

\newtheorem{theorem}{Theorem}[section]
\newtheorem{lemma}[theorem]{Lemma}

\theoremstyle{definition}

\theoremstyle{remark}

\begin{document}
\baselineskip =1.6\baselineskip

\begin{titlepage}
\title{\Large\bf  Asymptotic Behavior of the Newton-Boussinesq Equation
in a Two-Dimensional Channel } \vspace{20mm}

\author{ Guglielmo Fucci \vspace{3mm}\\
{\small  Department of  Physics}  \\
 {\small  New Mexico Institute of Mining and
Technology,  Socorro,  NM~87801, USA } \vspace{1cm}  \\
  Bixiang Wang
  \thanks {Supported in part by the Start-up Fund of
New Mexico Institute of Mining and Technology.},
  \quad   \quad  Preeti Singh \vspace{3mm}\\
  {\small  Department of  Mathematics }\\
 {\small New Mexico Institute of Mining and
Technology,  Socorro,  NM~87801, USA     }
}

\vspace{9mm}
\date{}
\end{titlepage}

\maketitle

\medskip

 \begin{abstract}
 We prove the existence of a global attractor
 for the Newton-Boussinesq equation defined
in a two-dimensional channel. The asymptotic compactness
of the equation is derived by the uniform estimates on the tails
of solutions. We also establish the regularity  of the global attractor.
\end{abstract}



\section{Introduction }
\setcounter{equation}{0}

In this paper, we investigate the asymptotic behavior of solutions of the
\nb defined in  an unbounded domain. Let $\Omega = (0, L) \times  \mathbb{R} $ where
$L$ is a positive number. Consider the system of equations defined in  $ (x,y) \in \Omega$
and $  t>0$:
\begin{equation}\label{1}
\partial_{t}\xi+u\partial_{x}\xi+v\partial_{y}\xi=\triangle\xi-
\frac{R_{a}}{P_{r}}\partial_{x}\theta+f(x,y),
\end{equation}
\begin{equation}\label{2}
 \triangle\Psi =\xi, \qquad u=\Psi_{y},\qquad v=-\Psi_{x},
\end{equation}
\begin{equation}\label{3}
\partial_{t}\theta+u\partial_{x}\theta+v\partial_{y}\theta=\frac{1}{P_{r}}\triangle\theta+g(x,y)\;,
\end{equation}
where  $\vec{u}=(u,v)$ is
the velocity vector of the fluid,  $\theta$  is  the flow
temperature,   $\Psi$ is the flow function, $\xi$ is
the vortex.  The positive  constants  $P_{r} $ and $R_{a} $ are
 the Prandtl number and the Rayleigh number,  respectively.
 The
external terms  $f$ and $g$ are  given in  $L^2(\Omega)$.

The \nb describes  many physical phenomena such as Benard flow, see,
\cite{che1, fei1} and the references therein.
If  the domain is bounded, the existence, uniqueness and
the asymptotic behavior of  solutions of  system \eqref{1}-\eqref{3} have been studied
by several authors, see, e.g.,  \cite{guo1, guo2, guo3, guo4}. In this paper we will examine the dynamical
behavior of  the  solutions when the system  is defined in the  unbounded two-dimensional channel
$\Omega$. More precisely, we will prove the  existence of a global attractor for  the system.
 Note that the unboundedness of the domain $\Omega$
introduces a major difficulty for proving the existence of a global attractor because
Sobolev embeddings are no longer compact in this case,  and hence the asymptotic compactness of the solution
operator cannot be obtained by a standard method.
Several approaches have been developed  to overcome this difficulty.
The energy equation method is one way to prove the asymptotic compactness
of equations  defined in   unbounded  domains. This idea   was first developed by Ball in
\cite{bal1, bal2} to deal with  the compactness of the wave equation and the Navier-Stokes
equation in bounded domains, and then
extended  by other authors  in \cite{ju1, moi1, ros1} to the Navier-Stokes equation in   unbounded domains.
Note that the energy equation of the Navier-Stokes equation
 in $L^2(\Omega)$ does not contain the nonlinear term.
 This fact together with the weak compactness can be used to prove the
 strong asymptotic compactness in $L^2(\Omega)$ (see, e.g., \cite{moi1, ros1}). However,
in our case, the energy
 equation for system \eqref{1}-\eqref{3} in $L^2(\Omega)$
 does contain the nonlinear term, and hence
 the energy equation approach does not apply. In this paper, we will employ
 the techniques  of uniform estimates on the tails of solutions to establish the asymptotic
 compactness of the Newton-Boussinesq equation.  This idea was develop in \cite{wan1}
 for proving the asymptotic compactness of the Reaction-Diffusion equation in unbounded
 domains, and   later used by several authors in \cite{ant1, ant2, arr1,  mor1, pri1, rod1, sun1}.

 This paper is organized  as follows. In the next section, we
 derive uniform estimates for the solutions of the system \eqref{1}-\eqref{3}
when $t \to \infty$,  which are necessary
 for proving  the existence of a bounded absorbing set and the asymptotic
 compactness of the equation.
 In Section 3, we first establish the asymptotic compactness of system \eqref{1}-\eqref{3}
by uniform estimates on the tails of solutions,  and then
 prove the existence of   a global attractor. The regularity of the global attractor is given in the last section.

In the sequel,  we adopt  the following notations.  The norm
of $L^2(\Omega)$  is  denoted by $||\cdot||$ which is
defined by mean of the usual inner product $(\cdot,\cdot)$. The
norm of  any Banach space $X$  is written as    $||\cdot||_{X}$.
In particular $||\cdot||_{p}$   represents  the norm  of
$L^{p}(\Omega)$.  The letter $C$
is a generic positive constant which may change its value from line to
line.

Throughout this paper,  we  frequently  use  the following inequality
\begin{equation}\label{5}
||u||_{4}<C||u||_{H^{1}(\Omega)}^{\frac{1}{2}}||u||^{\frac{1}{2}},\quad\forall
u\in H^{1}(\Omega),
\end{equation}
and the Poincare inequality
\begin{equation}\label{6}
  || u ||\leq  \lambda ||\nabla u
 ||\quad\forall  u \in H_{0}^{1}(\Omega),
\end{equation}
where $\lambda$ is a positive constant.

\section{Uniform Estimates of Solutions }
\setcounter{equation}{0}

 In this section, we  derive
   uniform estimates for the solutions of the system \eqref{1}-\eqref{3}
for large time. We also prove that the tails of solutions are uniformly small
when space and time variables are sufficiently large.

  Notice that system \eqref{1}-\eqref{3} can be rewritten as follows:
  for every $(x,y) \in \Omega$ and $t >0$,
\begin{equation}\label{7}
\frac{\partial\xi}{\partial
t}-\triangle\xi+J(\Psi,\xi)+\frac{R_{a}}{P_{r}}\frac{\partial\theta}{\partial
x}=f(x,y)\;,
\end{equation}
\begin{equation}\label{7_a1}
 \triangle \Psi =\xi \;,
\end{equation}
\begin{equation}\label{8}
\frac{\partial\theta}{\partial
t}-\frac{1}{P_{r}}\triangle\theta+J(\Psi,\theta)=g(x,y)\;,
\end{equation}
with  the   boundary conditions
\begin{equation}\label{10}
\xi|_{\partial\Omega}=0,\qquad\theta|_{\partial\Omega}=0,   \qquad \Psi|_{\partial \Omega} =0,
\end{equation}
and the initial conditions
\begin{equation}\label{11}
\xi(x,y,0)=\xi_{0}(x,y),\qquad\theta(x,y,0)=\theta_{0}(x,y),
\end{equation}
 where the functional $J$ is  given by
\begin{equation}\label{9}
J(u,v)=u_{y}v_{x}-u_{x}v_{y}\;.
\end{equation}
It is easy to verify that $J$ satisfies:
\begin{equation}\label{9a}
\int_{\Omega} J(u,v)v\;dxdy = 0 ,
\quad  {\mbox{ for all } } \ u \in H^1(\Omega), \ v \in H^2(\Omega) \cap H^1_0(\Omega ),
\end{equation}
\begin{equation}\label{one*}
 ||  J(u,v) || \le C \| u \|_{H^2} \| v \|_{H^2} ,
\quad  {\mbox{ for all } } \ u \in H^2(\Omega), \ v \in H^2(\Omega),
\end{equation}
\begin{equation}\label{two*}
 ||  J(u,v) || \le C \| u \|_{H^3} \|  \nabla v \| ,
\quad  {\mbox{ for all } } \ u \in H^3(\Omega), \ v \in H^1(\Omega).
\end{equation}

 It is standard to prove that problem (\ref{7})-(\ref{11}) is well
 posed in ${L}^{2}(\Omega)\times{}{L}^{2}(\Omega)$ (see, e.g.,
 \cite{guo1}). More precisely,
for every
$(\xi_{0},\theta_{0})\in  {L}^{2}(\Omega)\times{}{L}^{2}(\Omega)$,
  system  (\ref{7})-(\ref{11})
 has a unique solution $(\xi,\theta)$  such that for every $T>0$,
\begin{displaymath}
\xi\in
C^{0}\big([0,\infty),{}{L}^{2}(\Omega)\big)\cap{}{L}^{2}\big(0,T;H_{0}^{1}(\Omega)\big)\;,
\end{displaymath}
\begin{displaymath}
\theta\in
C^{0}\big([0,\infty),{}{L}^{2}(\Omega)\big)\cap{}{L}^{2}\big(0,T;H_{0}^{1}(\Omega)\big)\;.
\end{displaymath}
Therefore, we can define a semigroup
$\{S(t)\}_{t\ge 0}$ such that for every $ t \ge 0$, $S(t)$ maps
${L}^{2}(\Omega)\times{}{L}^{2}(\Omega)$ into itself
and $S(t)(\xi_0, \theta_0) =  (\xi (t), \theta(t))$.
We now start to derive uniform estimates for the dynamical system
$\{S(t)\}_{t\ge 0}$.

\begin{lemma}\label{lemma:1}
Suppose that
$(\xi_{0},\theta_{0})\in{}{L}^{2}(\Omega)\times{}{L}^{2}(\Omega)$.
 Then for every $T>0$, there is a constant $C>0$ such that the
solution $(\xi,\theta)$ of   system (\ref{7})-(\ref{11})  satisfies
\begin{displaymath}
||\xi(t)||+||\theta(t)||\leq C\qquad\forall t\in[0,T]\;,
\end{displaymath}
where $C$    depends   only on the data
$(\Omega,P_{r},R_{a})$, $T$ depends on the data
$(\Omega,P_{r},R_{a})$ and $R$ when $||\xi_{0}||\leq R$ and
$||\theta_{0}||\leq R$.
\end{lemma}

\begin{proof}
 We first consider equation (\ref{8}). By taking the
inner product of (\ref{8}) with $\theta$ in ${}{L}^{2}(\Omega)$
and using relation (\ref{9a}) we obtain
\begin{equation}\label{12}
\frac{1}{2}\frac{d}{dt}||\theta||^{2}+\frac{1}{P_{r}}||\nabla\theta||^{2}=\int_{\Omega}g\theta\;dxdy\;.
\end{equation}
By Poincair\'{e} inequality (\ref{6}) we find that
\begin{equation}\label{13}
\frac{1}{2}\frac{d}{dt}||\theta||^{2}+
+\frac{1}{2P_{r}}||\nabla\theta||^{2}
+
\frac{1}{2\lambda^{2}P_{r}}||\theta||^{2}\leq\int_{\Omega}g\theta\;dxdy\;.
\end{equation}
Notice that the right-hand side is bounded by
\begin{equation}\label{14}
\int_{\Omega}g\theta\;dxdy\leq||g||\;||\theta||\leq
C||g||^{2}+\frac{1}{4\lambda^{2}P_{r}}||\theta||^{2}\;.
\end{equation}
By  (\ref{12})-(\ref{14})  we find that
 \begin{equation}\label{14_aa1}
 \frac{d}{dt}||\theta||^{2}
+\frac{1}{P_{r}}||\nabla\theta||^{2}
+
\frac{1}{2 \lambda^{2}P_{r}}||\theta||^{2}\leq
 C, \quad\forall t\geq 0\;,
\end{equation}
which implies that
\begin{equation}\label{15}
\frac{d}{dt}||\theta||^{2}+C_{1}||\theta||^{2}\leq
C ,\quad\forall t\geq 0\;.
\end{equation}
Let $T>0$ be fixed and take $t \in [0, T]$, integrating \eqref{15}
over $(0, t)$
 we obtain
\begin{equation}\label{15a}
||\theta(t)||^{2}\leq C_{1}T+||\theta(0)||^{2}\leq
C_{1}T+R^{2}\leq C,\quad\forall t\in[0,T]\;.
\end{equation}
We  now consider  equation (\ref{7}). By taking the inner product
of (\ref{7}) with $\xi$ in ${}{L}^{2}(\Omega)$ and using relation
(\ref{9a}) we obtain
\begin{equation}\label{18}
\frac{1}{2}\frac{d}{dt}||\xi||^{2}+||\nabla\xi||^{2}+\frac{R_{a}}{P_{r}}\int_{\Omega}\theta_{x}\xi
dxdy=\int_{\Omega}f\xi dxdy\;.
\end{equation}
We notice that the following inequalities hold
\begin{equation}\label{17}
\frac{R_{a}}{P_{r}}\left|\int_{\Omega} \theta_x \xi
dxdy\right|=\frac{R_{a}}{P_{r}}\left|\int_{\Omega} \xi_x\theta
dxdy\right|\leq\frac{R_{a}}{P_{r}}||\nabla\xi||||\theta||\leq\frac{1}{4}||\nabla\xi||^{2}+C||\theta||^{2}\;,
\end{equation}
and
\begin{equation}\label{17a}
\left|\int_{\Omega}f\xi dxdy\right|\leq||f||||\xi||\leq
 \lambda ||f||||\nabla\xi||\leq\frac{1}{4}||\nabla\xi||^{2}+C\;.
\end{equation}
By (\ref{15a}), (\ref{17}) and (\ref{17a}), it follows from
(\ref{18}) that
\begin{equation}\label{19}
\frac{d}{dt}||\xi||^{2}+||\nabla\xi||^{2}\leq C,\quad\forall
t\in[0,T]\;.
\end{equation}
Then Poincair\`{e} inequality implies that
\begin{equation}\label{20}
\frac{d}{dt}||\xi||^{2}+C_{1}||\xi||^{2}\leq C,\quad\forall
t\in[0,T]\;.
\end{equation}
Integrating  (\ref{20})   on $(0, t)$ we obtain
\begin{equation}\label{20a}
||\xi(t)||^{2}\leq\ CT + \| \xi(0) \| \le C,\quad\forall t\in[0,T]\;.
\end{equation}
Combining   (\ref{15a}) and   (\ref{20a}) we
conclude that
\begin{equation}\label{20b}
||\theta(t)||+||\xi(t)||\leq C,\quad\forall t\in[0,T]\;.
\end{equation}
The proof is complete.
\end{proof}

\begin{lemma}\label{lemma:2}
Suppose that
$(\xi_{0},\theta_{0})\in{}{L}^{2}(\Omega)\times{}{L}^{2}(\Omega)$.
Then for the solution $(\xi,\theta)$ of   system
(\ref{7})-(\ref{11})  we have
\begin{displaymath}
||\xi(t)||+||\theta(t)||\leq M_{1}\qquad\forall t\geq t_{1}\;,
\end{displaymath}
and
\begin{displaymath}
\int_{t}^{t+1}||\nabla\xi(\tau)||^{2}d\tau+\int_{t}^{t+1}||\nabla\theta(\tau)||d\tau\leq
M_{2}\qquad\forall t\geq t_{1}\;,
\end{displaymath}
where $M_{1}$ and $M_{2}$ are constants depending only on the data
$(\Omega,P_{r},R_{a})$, $t_{1}$   depends  on the data
$(\Omega,P_{r},R_{a})$ and $R$ when $||\xi_{0}||\leq R$ and
$||\theta_{0}||\leq R$.
\end{lemma}

\begin{proof}
 By   (\ref{15})
and Gronwall inequality we infer that
\begin{equation}\label{16}
||\theta(t)||^{2}\leq e^{-C_{1}t}||\theta(0)||^{2}+C_{2}\leq
e^{-C_{1}t}R^{2}+C_{2}\leq 2C_{2}\quad\forall t\geq
t^{\ast}_{1}\;,
\end{equation}
where
$t^{\ast}_{1}=\frac{1}{C_{1}}\ln\left(\frac{R^{2}}{C_{2}}\right)$.
Moreover by (\ref{17}), (\ref{17a}) and (\ref{16}) we get from  (\ref{20}) that
\begin{equation}\label{16a}
\frac{d}{dt}||\xi||^{2}+C_{1}||\xi||^{2}\leq C,\quad\forall t\geq
t^{\ast}_{1}\;.
\end{equation}
By Lemma \ref{lemma:1} and  Gronwall inequality  we have
\begin{eqnarray}\label{21}
||\xi(t)||^{2}&\leq&
e^{-C_{1}(t-t^{\ast}_{1})}||\xi(t^{\ast}_{1})||^{2}+\frac{C}{C_{1}}\nonumber\\
&\leq&
e^{-C_{1}(t-t^{\ast}_{1})}C_{2}+\frac{C}{C_{1}}\leq\frac{2C}{C_{1}}\;,\quad\forall
t\geq
t^{\star}_{2}+\frac{1}{C_{1}}\ln\left(\frac{C_{1}C_{2}}{C}\right)\;,
\end{eqnarray}
Combining (\ref{16}) and (\ref{21}) we find that
\begin{equation}\label{23}
||\theta(t)||+||\xi(t)||\leq C ,\quad\forall t\geq t_{1}\;,
\end{equation}
where $t_{1}=\max\{t^{\ast}_{1},t^{\ast}_{2}\}$.
By \eqref{14_aa1} we obtain  that
\begin{equation}\label{24}
\frac{d}{d\tau}||\theta||^{2}+C||\nabla\theta||^{2}\leq
C_{1},\quad\forall \tau\geq t_{1}\;.
\end{equation}
Integrating
\eqref{24} on
$(t,t+1) $,  by     (\ref{23}) we have
  that
\begin{equation}\label{26}
\int_{t}^{t+1}||\nabla\theta(\tau)||^{2}d\tau\leq C , \quad\forall
t\geq t_{1}\;.
\end{equation}
By   (\ref{18}) and   (\ref{21}) we also
have
\begin{equation}\label{27}
\frac{d}{d\tau}||\xi||^{2}+C||\nabla\xi||^{2}\leq
C_{1},\quad\forall \tau\geq t_{1}\;.
\end{equation}
Integrating \eqref{27} on
$(t,t+1)$,   by \eqref{23} we get
\begin{equation}\label{28}
\int_{t}^{t+1}||\nabla\xi(\tau)||^{2}d\tau\leq C, \quad\forall
t\geq t_{1}\;.
\end{equation}
Then Lemma \ref{lemma:2} follows from \eqref{23}, \eqref{26} and \eqref{28}.
\end{proof}

We  now derive uniform estimates in $H^1(\Omega)$.

\begin{lemma}\label{lemma:3}
Suppose that
$(\xi_{0},\theta_{0})\in{}{L}^{2}(\Omega)\times{}{L}^{2}(\Omega)$.
Then for the solution $(\xi,\theta)$ of   system
(\ref{7})-(\ref{11}) we have
\begin{displaymath}
||\nabla\xi(t)||+||\nabla\theta(t)||\leq M_{3}\qquad\forall t\geq
t_{3}\;,
\end{displaymath}
and
\begin{displaymath}
\int_t^{t+1} (||\triangle \xi (t) ||^2 +||\triangle \theta (t) ||^2 )  \ dt \leq  M_3,\quad\forall t\geq t_3
 \;,
 \end{displaymath}
where $M_{3}$ is a constant depending only on the data
$(\Omega,P_{r},R_{a})$, $t_{3}$ depends on the data
$(\Omega,P_{r},R_{a})$ and $R$ when $||\xi_{0}||\leq R$ and
$||\theta_{0}||\leq R$.
\end{lemma}

\begin{proof}
Taking the inner
product of (\ref{7}) with $\triangle\xi$ in ${}{L}^{2}(\Omega)$
 we get
\begin{equation}\label{30}
\frac{1}{2}\frac{d}{dt}||\nabla\xi||^{2}+||\triangle\xi||^{2}=\int_{\Omega}J(\Psi,\xi)\triangle\xi
dxdy+\frac{R_{a}}{P_{r}}\int_{\Omega}\theta_{x}\triangle\xi
dxdy-\int_{\Omega}f\triangle\xi dxdy\;.
\end{equation}
Notice that the first term on the right-hand side of (\ref{30}) is
given by
\begin{equation}\label{30a}
\int_{\Omega}J(\Psi,\xi)\triangle\xi
dxdy=\int_{\Omega}\Psi_{y}\xi_{x}\triangle\xi
dxdy+\int_{\Omega}\Psi_{x}\xi_{y}\triangle\xi dxdy\;.
\end{equation}
We  now estimate the first term on the right-hand side of
(\ref{30a}). By  \eqref{5} and Lemma \ref{lemma:2}  we have the following
estimates for $t \ge T$,
\begin{eqnarray}\label{31}
\int_{\Omega}\Psi_{y}\xi_{x}\triangle\xi
dxdy\leq||\Psi_{y}||_{4}||\xi_{x}||_{4}||\triangle\xi||\leq
C||\Psi_{y}||^{\frac{1}{2}}||\Psi_{y}||^{\frac{1}{2}}_{H^{1}}||\xi_{x}||^{\frac{1}{2}}||\xi_{x}||^{\frac{1}{2}}_{H^{1}}||\triangle\xi||\nonumber\\
\leq
C||\Psi||_{H^{2}}||\nabla\xi||^{\frac{1}{2}}||\triangle\xi||^{\frac{3}{2}}\leq
C||\nabla\xi||^{\frac{1}{2}}||\triangle\xi||^{\frac{3}{2}}\leq\frac{1}{8}||\triangle\xi||^{2}+C||\nabla\xi||^{2}\;.
\end{eqnarray}
Similarly for the second  term on the right-hand side of \eqref{30a}   we have
\begin{equation}\label{32}
\int_{\Omega}\Psi_{x}\xi_{y}\triangle\xi
dxdy\leq\frac{1}{8}||\triangle\xi||^{2}+C||\nabla\xi||^{2}\;.
\end{equation}
It follows from (\ref{30a}) and (\ref{32}) that
\begin{equation}\label{33}
\int_{\Omega}J(\Psi,\xi)\triangle\xi
dxdy\leq\frac{1}{4}||\triangle\xi||^{2}+C||\nabla\xi||^{2}\;.
\end{equation}
Note that the last  two  terms on the right-hand side of
\eqref{30} are bounded by \be \label{33_aa1} C \left|
\int_{\Omega}\theta_{x}\triangle\xi dxdy \right| +\left|
\int_{\Omega}f\triangle\xi dxdy\right| \leq
\frac{1}{4}||\triangle\xi||^{2}+C||\nabla\theta||^{2} +C. \ee From
(\ref{30}) and  (\ref{33})-(\ref{33_aa1}) we have
\begin{equation}\label{34}
\frac{d}{dt}||\nabla\xi||^{2}+||\triangle\xi||^{2}\leq
C\left(||\nabla\xi||^{2}+||\nabla\theta||^{2}\right)+C,\quad\forall
t\geq T\;.
\end{equation}
Taking the inner product of
(\ref{8}) with $\triangle\theta$   we get
\begin{equation}\label{35}
\frac{1}{2}\frac{d}{dt}||\nabla\theta||^{2}+\frac{1}{P_{r}}||\triangle\theta||^{2}=\int_{\Omega}J(\Psi,\theta)\triangle\theta
dxdy+\int_{\Omega}g\triangle\theta dxdy\;.
\end{equation}
By arguments similar to \eqref{33} and \eqref{33_aa1},  we   obtain   that
\begin{equation}\label{37}
\frac{d}{dt}||\nabla\theta||^{2}+\frac{1}{2P_{r}}||\triangle\theta||^{2}\leq
C\left(||\nabla\xi||^{2}+||\nabla\theta||^{2}\right)+C,\quad\forall
t\geq T\;.
\end{equation}
Let $\alpha=\min\left\{1,\frac{1}{2P_{r}}\right\}$.
 Then  from (\ref{34}) and (\ref{37}),  we have
\begin{equation}\label{38}
\frac{d}{dt}\left(||\nabla\xi||^{2}+||\nabla\theta||^{2}\right)+\alpha\left(||\triangle\theta||^{2}+||\triangle\xi||^{2}\right)\leq
C\left(||\nabla\xi||^{2}+||\nabla\theta||^{2}\right)+C,\quad\forall
t\geq T\;.
\end{equation}
By  the uniform Gronwall inequality and Lemma  \ref{lemma:2},
we find  from \eqref{38} that
\begin{equation}\label{39}
||\nabla\xi(t) ||+||\nabla\theta (t) ||\leq  C,\quad\forall t\geq
T+1\;.
\end{equation}
Integrating  \eqref{38} on $(t, t+1)$, by    \eqref{39}  we  get
\begin{equation}
\label{39_aa1}
\int_t^{t+1} (||\triangle \xi (t) ||^2 +||\triangle \theta (t) ||^2 ) dt \leq  C,\quad\forall t\geq
T+1\;.
\end{equation}
Then Lemma \ref{lemma:3} follows from \eqref{39}-\eqref{39_aa1}.
\end{proof}

Next we establish the uniform estimates on the tails of solutions which are crucial
for proving the asymptotic  compactness of the solution operator.
Given $k>0$, we denote by $\Omega_{k}$ the set
$\Omega_{k}=\{(x,y)\in\Omega:\ |y|\leq k\}$ and
$\Omega\backslash\Omega_{k}$ the complement of $\Omega_{k}$. For
our purpose, we choose a smooth cut-off function $\phi$ such that
$0\leq\phi(s)\leq 1$ and
\begin{equation}\label{40}
\left\{ \begin{array}{ll} \phi(s)=0 & \textrm{if $|s|<1$}\\
\phi(s)=1 & \textrm{if $|s|>2$}\;.\\
\end{array}\right.
\end{equation}
Then, we have the following Poincair\'{e} type of inequality

\begin{lemma}\label{lemma:4}
Let $v\in H_{0}^{1}(\Omega)$ and $\phi$ be given as above. Then
$\exists\;\alpha>0$ and $\beta>0$ such that $\forall k>0$:
\begin{displaymath}
\int_{\Omega}\phi^{2}\left(\frac{|y|^{2}}{k^{2}}\right)|\nabla
v|^{2}dxdy\geq\alpha\int_{\Omega}\phi^{2}\left(\frac{|y|^{2}}{k^{2}}\right)v^{2}dxdy-\frac{\beta}{k^{2}}\int_{\Omega}v^{2}dxdy\;.
\end{displaymath}
\end{lemma}

\begin{proof}
 By Poincair\`{e} inequality (\ref{6}), we have
\begin{equation}\label{41}
 \int_{\Omega} \phi^{2}  \left(\frac{|y|^{2}}{k^{2}}\right) v^{2}dxdy\leq
\lambda^{2}\int_{\Omega}\left|\nabla\left ( \phi\left(\frac{|y|^{2}}{k^{2}}\right)
v\right)\right|^{2}dxdy\;,
\end{equation}
Notice that
$$
\int_{\Omega}\left|\nabla\left( \phi\left(\frac{|y|^{2}}{k^{2}}\right)
v\right) \right|^{2}dxdy
\leq \int_{\Omega}\phi^{2}\left(\frac{|y|^{2}}{k^{2}}\right)|\nabla
v|^{2}dxdy+4\int_{\Omega}  \left (\phi^\prime \left(\frac{|y|^{2}}{k^{2}}\right) \right ) ^{2}
 \  \frac{y^{2}}{k^{4}}\; v ^{2}dxdy
 $$
 $$
\leq \int_{\Omega}\phi^{2}\left(\frac{|y|^{2}}{k^{2}}\right)|\nabla
v|^{2}dxdy+4\int_{k \le |y| \le \sqrt{2} k}  \left (\phi^\prime \left(\frac{|y|^{2}}{k^{2}}\right) \right ) ^{2}
 \  \frac{y^{2}}{k^{4}}\; v ^{2}dxdy
 $$
 $$
 \leq  \int_{\Omega}\phi^{2}\left(\frac{|y|^{2}}{k^{2}}\right)|\nabla
v|^{2}dxdy+\frac{C}{k^{2}}\int_{k\leq|y|\leq\sqrt{2}k}|v|^{2}dxdy
$$
\begin{eqnarray}\label{42}
  \leq  \int_{\Omega}\phi^{2}\left(\frac{|y|^{2}}{k^{2}}\right)|\nabla
v|^{2}dxdy+
\frac{C}{k^{2}}
\int_{\Omega}|v|^{2}dxdy\;.
\end{eqnarray}
From (\ref{41}) and (\ref{42}) it  follows that
\begin{equation}\label{43}
\int_{\Omega} \phi ^2 \left(\frac{|y|^{2}}{k^{2}}\right)   v ^{2}dxdy\leq
\lambda ^{2}\int_{\Omega}\phi^{2}\left(\frac{|y|^{2}}{k^{2}}\right)|\nabla
v|^{2}dxdy +\frac{C }{k^{2}}\int_{\Omega}|v|^{2}dxdy\;,
\end{equation}
which implies Lemma \ref{lemma:4}.  The proof is complete.
\end{proof}

\begin{lemma}\label{lemma:5}
Given  $\epsilon>0$, then there exist $t_3>0$ and $k_0>0$ such that
 the solution  $(\xi,\theta)$   of system
(\ref{7})-(\ref{11}) with the initial condition
$(\xi_{0},\theta_{0})$
satisfies
\begin{displaymath}
\int_{\Omega\setminus\Omega_{k_{0}}}\left(|\xi (t) |^{2}+|\theta (t) |^{2}\right)dxdy\leq\epsilon,\quad\forall
t\geq t_{3}\;,
\end{displaymath}
where $k_{0}$ depends only on the data $(\Omega,P_{r},R_{a})$ and
$\epsilon$, $t_{3}$ depends only on $(\Omega,P_{r},R_{a})$,
$\epsilon$ and $R$ when $||\xi_{0}||\leq R$ and
$||\theta_{0}||\leq R$.
\end{lemma}

\begin{proof}
 Multiplying   (\ref{8}) by
  $\phi^{2}\left(\frac{|y|^{2}}{k^{2}}\right)\theta(x,y,t)$
and then integrating the resulting identity over $\Omega$, we obtain
\begin{eqnarray}\label{44}
\frac{1}{2}\frac{d}{dt}\int_{\Omega}|\theta|^{2}\phi^{2}\left(\frac{|y|^{2}}{k^{2}}\right)dxdy-\frac{1}{P_{r}}\int_{\Omega} (\theta\triangle\theta) \phi^{2}\left(\frac{|y|^{2}}{k^{2}}\right)dxdy \nonumber\\
=\int_{\Omega}g\theta\phi^{2}\left(\frac{|y|^{2}}{k^{2}}\right)dxdy
-\int_{\Omega}J(\Psi,\theta)\theta\phi^{2}\left(\frac{|y|^{2}}{k^{2}}\right)dxdy. 
\end{eqnarray}
We now  estimate every term in (\ref{44}). We first have, by  Lemma \ref{lemma:4}, 
$$
-\frac{1}{P_{r}}\int_{\Omega} (\theta\triangle\theta)  \phi^{2}\left(\frac{|y|^{2}}{k^{2}}\right)dxdy  
$$
$$
=\frac{1}{P_{r}}\int_{\Omega}\phi^{2}\left(\frac{|y|^{2}}{k^{2}}\right)|\nabla \theta|^{2}dxdy
+\frac{4}{P_{r}}\int_{\Omega}\phi\left(\frac{|y|^{2}}{k^{2}}\right)
\phi^{\prime}\left(\frac{|y|^{2}}{k^{2}}\right)\theta\theta_{y}\frac{y}{k^{2}}\;dxdy
$$
\begin{eqnarray}\label{45}
\geq\frac{\alpha}{P_{r}}\int_{\Omega}\phi^{2}\left(\frac{|y|^{2}}{k^{2}}\right)|\theta|^{2}dxdy-\frac{C}{k^{2}}||\theta||^{2} 
+\frac{4}{P_{r}}\int_{\Omega}\phi\left(\frac{|y|^{2}}{k^{2}}
\right)\phi^{\prime}\left(\frac{|y|^{2}}{k^{2}}\right)\theta\theta_{y}\frac{y}{k^{2}}\;dxdy .
\end{eqnarray}
For the last term on the right-hand side of (\ref{44}) we obtain,
by integration by parts,
$$
\int_{\Omega}J(\Psi,\theta)\theta\phi^{2}\left(\frac{|y|^{2}}{k^{2}}\right)dxdy 
$$
$$
=\int_{\Omega}\Psi_{y}\theta_{x}\theta\phi^{2}\left(\frac{|y|^{2}}{k^{2}}\right)dxdy-\int_{\Omega}\Psi_{x}\theta_{y}\theta\phi^{2}\left(\frac{|y|^{2}}{k^{2}}\right)dxdy
$$
$$
=\int_{\Omega}\Psi_{y}\left(\frac{1}{2}\;\theta^{2}\right)_{x}\phi^{2}\left(\frac{|y|^{2}}{k^{2}}\right)dxdy-\int_{\Omega}\Psi_{x}\left(\frac{1}{2}\;\theta^{2}\right)_{y}\phi^{2}\left(\frac{|y|^{2}}{k^{2}}\right)dxdy
$$
$$
=-\frac{1}{2}\int_{\Omega}\Psi_{yx}\theta^{2}\phi^{2}\left(\frac{|y|^{2}}{k^{2}}\right)dxdy
+\frac{1}{2}\int_{\Omega}\Psi_{yx}\theta^{2}\phi^{2}\left(\frac{|y|^{2}}{k^{2}}\right)dxdy
$$
$$
+2\int_{\Omega}\Psi_{x}\theta^{2}\phi^{\prime}\left(\frac{|y|^{2}}{k^{2}}\right)\phi\left(\frac{|y|^{2}}{k^{2}}
\right)\frac{y}{k^{2}}dxdy
$$
\begin{eqnarray}\label{45a}
=2\int_{\Omega}\Psi_{x}\theta^{2}\phi^{\prime}\left(\frac{|y|^{2}}{k^{2}}\right)\phi\left(\frac{|y|^{2}}{k^{2}}\right)\frac{y}{k^{2}}dxdy.
\end{eqnarray}
It follows, from (\ref{44}) through (\ref{45a}) that
$$
\phantom{,}\frac{1}{2}\frac{d}{dt}\int_{\Omega}|\theta|^{2}
\phi^{2}\left(\frac{|y|^{2}}{k^{2}}\right)dxdy+\frac{\alpha}{P_{r}}
\int_{\Omega}\phi^{2}\left(\frac{|y|^{2}}{k^{2}}\right)|\theta|^{2}dxdy
$$
$$
=\int_{\Omega}g\theta\phi^{2}\left(\frac{|y|^{2}}{k^{2}}\right)dxdy
 -\frac{4}{P_{r}}\int_{\Omega}\phi\left(\frac{|y|^{2}}{k^{2}}\right)\phi^{\prime}\left(\frac{|y|^{2}}{k^{2}}\right)\theta\theta_{y}\frac{y}{k^{2}}\;dxdy
 $$
\begin{eqnarray}\label{46}
- 2\int_{\Omega}\Psi_{x}\theta^{2}\phi^{\prime}\left(\frac{|y|^{2}}{k^{2}}\right)
\phi\left(\frac{|y|^{2}}{k^{2}}\right)\frac{y}{k^{2}}dxd
+\frac{C}{k^{2}}||\theta||^{2}\;.
\end{eqnarray}
Note that the first term on the right-hand side of (\ref{46}) is
bounded by
\begin{eqnarray}\label{47}
\left|\int_{\Omega}g\theta\phi^{2}\left(\frac{|y|^{2}}{k^{2}}\right)dxdy\right|=\left|\int_{|y|\geq
k}g\theta\phi^{2}\left(\frac{|y|^{2}}{k^{2}}\right)dxdy\right|    \nonumber\\
\leq\left(\int_{|y|\geq
k}g^{2}dxdy\right)^{\frac{1}{2}}\left(\int_{|y|\geq
k}\phi^{4}\left(\frac{|y|^{2}}{k^{2}}\right)\theta^{2}
dxdy\right)^{\frac{1}{2}}  \nonumber\\
\leq\left(\int_{|y|\geq
k}g^{2}dxdy\right)^{\frac{1}{2}}\left(\int_{\Omega}\phi^{2}\left(\frac{|y|^{2}}{k^{2}}\right)\theta^{2}
dxdy\right)^{\frac{1}{2}}  \nonumber\\
\leq C\int_{|y|\geq
k}|g|^{2}dxdy+\frac{\alpha}{2P_{r}}\int_{\Omega}|\theta|^{2}\phi^{2}\left(\frac{|y|^{2}}{k^{2}}\right)dxdy\;.
\end{eqnarray}
For the second term on the right-hand side of (\ref{46}) we have
$$
\frac{4}{P_{r}}\left|\int_{\Omega}\phi\left(\frac{|y|^{2}}{k^{2}}\right)\phi^{\prime}\left(\frac{|y|^{2}}{k^{2}}\right)\theta\theta_{y}\frac{y}{k^{2}}\;dxdy\right|
$$
$$
=\frac{4}{P_{r}}\left|\int_{k\leq|y|\leq\sqrt{2k}}\phi\left(\frac{|y|^{2}}{k^{2}}\right)\phi^{\prime}\left(\frac{|y|^{2}}{k^{2}}\right)\theta\theta_{y}\frac{y}{k^{2}}\;dxdy\right|
$$
\begin{eqnarray}\label{48}
\leq\frac{C}{k}\int_{k\leq|y|\leq\sqrt{2k}}|\theta||\theta_{y}|dxdy
\leq\frac{C}{k}||\theta||||\nabla\theta||\leq\frac{C}{k}\;.
\end{eqnarray}
where the last inequality is obtained by Lemmas (\ref{lemma:2}) and
(\ref{lemma:3}). The third term on the right-hand side is bounded
by
$$
2\left|\int_{\Omega}\Psi_{x}\theta^{2}\phi^{\prime}
\left(\frac{|y|^{2}}{k^{2}}\right)\phi\left(\frac{|y|^{2}}{k^{2}}\right)\frac{y}{k^{2}}dxdy\right|
$$
$$
=2\left|\int_{k\leq|y|\leq\sqrt{2k}}\Psi_{x}\theta^{2}\phi^{\prime}
\left(\frac{|y|^{2}}{k^{2}}\right)\phi\left(\frac{|y|^{2}}{k^{2}}\right)\frac{y}{k^{2}}dxdy\right|
$$
$$
\leq \frac{C}{k}\int_{k\leq|y|\leq\sqrt{2k}}|\Psi_{x}||\theta|^{2}dxdy
\leq\frac{C}{k}\int_{\Omega}|\Psi_{x}||\theta|^{2}dxdy
$$
\begin{eqnarray}\label{49}
\leq\frac{C}{k}||\Psi_x ||_{6}||\theta||_{3}||\theta||
\leq \frac{C}{k}||\Psi||_{H^{2}}||\theta||_{H^{1}}||\theta||\leq\frac{C}{k}\;.
\end{eqnarray}
It follows, from (\ref{46}) through (\ref{49}) that for $k\geq 1$
\begin{equation}\label{49a}
\frac{d}{dt}\int_{\Omega}|\theta|^{2}\phi^{2}\left(\frac{|y|^{2}}{k^{2}}\right)dxdy+\frac{\alpha}{P_{r}}\int_{\Omega}|\theta|^{2}\phi^{2}\left(\frac{|y|^{2}}{k^{2}}\right)dxdy\leq\frac{C}{k}+C_{1}\int_{|y|\geq
k}|g|^{2}dxdy\;.
\end{equation}
Now, since $g\in{}{L}^{2}(\Omega)$, given $\epsilon>0$, there
exists $k_{1}>0$ such that
\begin{equation}\label{50}
C_{1}\int_{|y|\geq
k}|g|^{2}dxdy\leq\frac{\epsilon}{2},\quad\forall k\geq
k_{1}(\epsilon)\;.
\end{equation}
Let $k_{2}=\max\{k_{1},\frac{2C}{\epsilon}\}$, then by (\ref{49a})
and (\ref{50}) we obtain that  for all  $k\geq k_{2}$ and  $ t\geq T,$
\begin{equation}\label{51}
\frac{d}{dt}\int_{\Omega}|\theta|^{2}\phi^{2}\left(\frac{|y|^{2}}{k^{2}}\right)dxdy+\frac{\alpha}{P_{r}}\int_{\Omega}|\theta|^{2}\phi^{2}\left(\frac{|y|^{2}}{k^{2}}\right)dxdy\leq\epsilon.
\end{equation}
Applying  Gronwall  lemma to (\ref{51}), by
Lemma \ref{lemma:2} we find that,   for all   $ k\geq k_{2}$,
\begin{eqnarray}\label{52}
\int_{\Omega}\phi^{2}\left(\frac{|y|^{2}}{k^{2}}\right)|\theta|^{2}dxdy\leq
e^{-\frac{\alpha}{P_{r}}(t-T)}\int_{\Omega}\phi^{2}\left(\frac{|y|^{2}}{k^{2}}\right)|\theta(T)|^{2}dxdy
+ \frac{\epsilon\alpha}{P_{r}}\nonumber\\
\leq
e^{-\frac{\alpha}{P_{r}}(t-T)}||\theta(T)||^{2}+\frac{\epsilon\alpha}{P_{r}}\leq\frac{2\alpha\epsilon}{P_{r}},
\end{eqnarray}
 for all $ t\geq
T_{1}=T-\frac{P_{r}}{\alpha}\ln\left(\frac{\alpha\epsilon}{CP_{r}}\right)$.
We now estimate
$\int_{\Omega}\phi^{2}\left(\frac{|y|^{2}}{k^{2}}\right)|\xi(x,y,t)|^{2}dxdy$.
Multiplying (\ref{7}) by
$\phi^{2}\left(\frac{|y|^{2}}{k^{2}}\right)\xi(x,y,t)$ and then
integrating by parts we get
$$
 \frac{1}{2}\frac{d}{dt}\int_{\Omega}|\xi|^{2}\phi^{2}\left(\frac{|y|^{2}}{k^{2}}\right)dxdy-\int_{\Omega} ( \xi\triangle\xi)  \phi^{2}\left(\frac{|y|^{2}}{k^{2}}\right)dxdy+\int_{\Omega}J(\Psi,\xi)\xi\phi^{2}\left(\frac{|y|^{2}}{k^{2}}\right)dxdy\nonumber\\
$$
\begin{eqnarray}\label{53}
=\int_{\Omega}f\xi\phi^{2}\left(\frac{|y|^{2}}{k^{2}}\right)dxdy
+\frac{R_{a}}{P_{r}}\int_{\Omega} \theta_x  \xi\phi^{2}\left(\frac{|y|^{2}}{k^{2}}\right)dxdy\;.
\end{eqnarray}
For the second term on the left-hand side of (\ref{53}), by Lemma
\ref{lemma:4} we have
$$
-\int_{\Omega} ( \xi\triangle\xi) \phi^{2}\left(\frac{|y|^{2}}{k^{2}}\right)dxdy
$$
$$
=\int_{\Omega}|\nabla\xi|^{2}\phi^{2}\left(\frac{|y|^{2}}{k^{2}}\right)dxdy+4\int_{\Omega}\xi\xi_{y}\phi\left(\frac{|y|^{2}}{k^{2}}\right)\phi^{\prime}\left(\frac{|y|^{2}}{k^{2}}\right)\frac{y}{k^{2}}\;dxdy
$$
$$
\geq \frac{1}{2}\int_{\Omega}\phi^{2}\left(\frac{|y|^{2}}{k^{2}}\right)|\nabla\xi|^{2}dxdy+\frac{\alpha}{2}\int_{\Omega}\phi^{2}\left(\frac{|y|^{2}}{k^{2}}\right)|\xi|^{2}dxdy
$$
\begin{eqnarray}\label{54}
-\frac{C}{k^{2}}||\xi||^{2}
+ 4\int_{\Omega}\xi\xi_{y}\phi\left(\frac{|y|^{2}}{k^{2}}\right)\phi^{\prime}\left(\frac{|y|^{2}}{k^{2}}\right)\frac{y}{k^{2}}\;dxdy\;.
\end{eqnarray}
By (\ref{53}) and (\ref{54}) we find that the following inequality
holds
\begin{eqnarray}\label{55}
\frac{1}{2}\frac{d}{dt}\int_{\Omega}|\xi|^{2}\phi^{2}\left(\frac{|y|^{2}}{k^{2}}\right)dxdy+\frac{1}{2}\int_{\Omega}|\nabla\xi|^{2}\phi^{2}\left(\frac{|y|^{2}}{k^{2}}\right)dxdy+\frac{\alpha}{2}\int_{\Omega}\phi^{2}\left(\frac{|y|^{2}}{k^{2}}\right)|\xi|^{2}dxdy
\nonumber\\
\leq\int_{\Omega}f\xi\phi^{2}\left(\frac{|y|^{2}}{k^{2}}\right)dxdy-\int_{\Omega}J(\Psi,\xi)\xi\phi^{2}\left(\frac{|y|^{2}}{k^{2}}\right)dxdy+\frac{R_{a}}{P_{r}}\int_{\Omega}\theta_{x}\xi\phi^{2}\left(\frac{|y|^{2}}{k^{2}}\right)dxdy\nonumber\\
-4\int_{\Omega}\xi\xi_{y}\phi\left(\frac{|y|^{2}}{k^{2}}\right)\phi^{\prime}\left(\frac{|y|^{2}}{k^{2}}\right)\frac{y}{k^{2}}\;dxdy+\frac{C}{k^{2}}||\xi||^{2}\;.\qquad\qquad
\end{eqnarray}
Note that the third term on the right-hand side of (\ref{55}) is
bounded by
$$
\frac{R_{a}}{P_{r}}\left|\int_{\Omega}\theta_{x}\xi\phi^{2}\left(\frac{|y|^{2}}{k^{2}}\right)dxdy\right|\leq\frac{R_{a}}{P_{r}}\left|\int_{\Omega}\theta\xi_{x}\phi^{2}\left(\frac{|y|^{2}}{k^{2}}\right)dxdy\right|
$$
$$
\leq\frac{1}{2}\int_{\Omega}\xi_{x}^{2}\phi^{2}\left(\frac{|y|^{2}}{k^{2}}\right)dxdy
+\frac{R^{2}_{a}}{2P^{2}_{r}}\int_{\Omega}|\theta|^{2}\phi^{2}\left(\frac{|y|^{2}}{k^{2}}\right)
$$
\begin{eqnarray}\label{56}
\leq\frac{1}{2}\int_{\Omega}|\nabla\xi|^{2}\phi^{2}\left(\frac{|y|^{2}}{k^{2}}\right)dxdy+C\epsilon\;,\qquad\qquad
\end{eqnarray}
where the last inequality is obtained by (\ref{52}). It follows,
then, from (\ref{55}) and (\ref{56}) that
$$
\frac{1}{2}\frac{d}{dt}\int_{\Omega}|\xi|^{2}\phi^{2}\left(\frac{|y|^{2}}{k^{2}}\right)dxdy
+\frac{\alpha}{2}\int_{\Omega}|\xi|^{2}\phi^{2}\left(\frac{|y|^{2}}{k^{2}}\right)dxdy
$$
$$
\leq\int_{\Omega}f\xi\phi^{2}\left(\frac{|y|^{2}}{k^{2}}\right)dxdy
-\int_{\Omega}J(\Psi,\xi)\xi\phi^{2}\left(\frac{|y|^{2}}{k^{2}}\right)dxdy
$$
\begin{eqnarray}\label{56a}
-4\int_{\Omega}\xi\xi_{y}\phi\left(\frac{|y|^{2}}{k^{2}}\right)\phi^{\prime}\left(\frac{|y|^{2}}{k^{2}}\right)\frac{y}{k^{2}}\;dxdy+\frac{C}{k^{2}}||\xi||^{2}+C_{1}\epsilon\;.\quad
\end{eqnarray}
By similar arguments used in (\ref{46}), after detailed
calculations, we find that for $k\geq k_{2}$ and $t\geq T_{1}$,  the
right-hand side of (\ref{56a}) is bounded by
\begin{displaymath}
C\epsilon+\frac{C_{1}}{k}+C_{2}\int_{|y|\geq
k}|f|^{2}dxdy+\frac{\alpha}{4}\int_{\Omega}|\xi|^{2}\phi^{2}\left(\frac{|y|^{2}}{k^{2}}\right)dxdy\;,
\end{displaymath}
and hence there is $k_{3}>0$ such that  for all   $k\geq k_{3}$ and
$t\geq T_{1}$,
\begin{displaymath}
\frac{d}{dt}\int_{\Omega}|\xi|^{2}\phi^{2}\left(\frac{|y|^{2}}{k^{2}}\right)dxdy
+\frac{\alpha}{2}\int_{\Omega}|\xi|^{2}\phi^{2}\left(\frac{|y|^{2}}{k^{2}}\right)dxdy\leq
C\epsilon\;.
\end{displaymath}
By Gronwall lemma, we find that for any $k\geq k_{3}$,
\begin{eqnarray}\label{56b}
\int_{\Omega}|\xi|^{2}\phi^{2}\left(\frac{|y|^{2}}{k^{2}}\right)dxdy \leq
e^{-\frac{\alpha}{2}(t-T_{1})}\int_{\Omega}|\xi(T_{1})|^{2}\phi^{2}\left(\frac{|y|^{2}}{k^{2}}\right)dxdy+C\epsilon\nonumber\\
\leq  e^{-\frac{\alpha}{2}(t-T_{1})}||\xi(T_{1})||^{2}+C\epsilon\leq
C_{1}e^{-\frac{\alpha}{2}(t-T_{1})}+C\epsilon\leq
2C\epsilon\;,\quad
\end{eqnarray}
for any $t\geq
T_{2}=T_{1}-\frac{2}{\alpha}\ln\frac{C\epsilon}{C_{1}}$.
By (\ref{52}) and (\ref{56b}) we see that for any $k\geq k_{3}$
and $t\geq T_{2}$,
\begin{equation}\label{57}
\int_{\Omega}\phi^{2}\left(\frac{|y|^{2}}{k^{2}}\right)\left(|\theta|^{2}+|\xi|^{2}\right)dxdy \le C\epsilon\;,
\end{equation}
and hence for all $k\geq k_{3}$ and $t\geq T_{1}$,
\begin{equation}\label{58}
\int_{|y|\geq\sqrt{2k}}\left(|\theta|^{2}+|\xi|^{2}\right)dxdy\leq\int_{\Omega}\phi^{2}\left(\frac{|y|^{2}}{k^{2}}\right)\left(|\theta|^{2}+|\xi|^{2}\right)dxdy \le C\epsilon\;,
\end{equation}
which implies Lemma \ref{lemma:5}. The proof is complete.
\end{proof}

\section{Existence of Global Attractors}
\setcounter{equation}{0}
In this section,  we prove the existence of global attractors for
  problem (\ref{7})-(\ref{8}) in
${}{L}^{2}(\Omega)\times{}{L}^{2}(\Omega)$. To this end, we need
to establish the asymptotic compactness of the solution operator
which is stated as follows.
\begin{lemma}\label{lemma:6}
The dynamical system $\{S(t)\}_{t\geq0}$ is asymptotically compact
in ${}{L}^{2}(\Omega)\times{}{L}^{2}(\Omega)$, i.e., if
$t_{n}\rightarrow\infty$ and
$\{(\xi_{0,n},\theta_{0,n})\}_{n=1}^{\infty}$ is bounded in
${}{L}^{2}(\Omega)\times{}{L}^{2}(\Omega)$, then the sequence
$\{S(t_{n})(\xi_{0,n},\theta_{0,n})\}_{n=1}^{\infty}$ has a
convergent subsequence.
\end{lemma}

\begin{proof}
 Since $\{(\xi_{0,n},\theta_{0,n})\}_{n=1}^{\infty}$
is bounded in ${}{L}^{2}(\Omega)\times{}{L}^{2}(\Omega)$, there is
$R>0$ such that
\begin{equation}\label{59}
||\xi_{0,n}||+||\theta_{0,n}||\leq R,\quad\forall
n\in\mathbb{Z}^{+}\;.
\end{equation}
By Lemma \ref{lemma:3}, there is a positive number $M$,
depending on $(\Omega,P_{r},R_{a})$, such that for every
$(\xi_{0},\theta_{0})\in{}{L}^{2}(\Omega)\times{}{L}^{2}(\Omega)$
with $||\xi_{0}||+||\theta_{0}||\leq R$, the following holds
\begin{equation}\label{60}
||S(t)(\xi_{0},\theta_{0})||_{H^{1}_0(\Omega)\times
H^{1}_0(\Omega)}\leq M,\quad\forall t\geq T_{1}\;,
\end{equation}
where $T_{1}$ depends on $(\Omega,P_{r},R_{a})$ and $R$. Since
$t_{n}\rightarrow\infty$, there is $N_1>0$ such that $t_{n}\geq
T_{1}$ for all $n\geq N_1$. Therefore we have, for $n\geq N_1$,
\begin{equation}\label{61}
||S(t_n)(\xi_{0,n},\theta_{0,n})||_{H^{1}_0(\Omega)\times
H^{1}_0(\Omega)}\leq M\;.
\end{equation}
By (\ref{61}) we find that there is $(\xi,\theta)\in
H^{1}_0(\Omega)\times H^{1}_0(\Omega)$ such that, up to a subsequence,
\begin{equation}\label{62}
S(t_{n})(\xi_{0,n},\theta_{0,n})\rightharpoonup(\xi,\theta)\quad\textrm{in}\quad{}{L}^{2}
(\Omega)\times{}{L}^{2}(\Omega)\quad\textrm{and}\quad
H^{1}_0(\Omega)\times H^{1}_0(\Omega)\;.
\end{equation}
Given $\epsilon>0$, by Lemma \ref{lemma:5}, there are positive
numbers $k_{1}$ and $T_{2}$ such that for any $k\geq k_{1}$ and
$t\geq T_{2}$, $S(t)(\xi_{0},\theta_{0})$, with
$||(\xi_{0},\theta_{0})||\leq R$, satisfies
\begin{equation}\label{63}
\int_{\Omega\setminus\Omega_{k}}\left(|S(t)\xi_{0}|^{2}+|S(t)\theta_{0}|^{2}\right)dxdy\leq\frac{\epsilon}{5}\;.
\end{equation}
Let $N_{2}$ be large enough such that $t_{n}\geq T_{2}$ for all
$n\geq N_{2}$. Then by (\ref{63}) we obtain, for $n\geq N_{2}$,
\begin{equation}\label{64}
\int_{\Omega\setminus\Omega_{k}}\left(|S(t_n)\xi_{0,n}|^{2}+|S(t_n)\theta_{0,n}|^{2}\right)dxdy\leq\frac{\epsilon}{5}\;.
\end{equation}
Notice that (\ref{61}) implies that the sequence
$\{S(t_{n})(\xi_{0,n},\theta_{0,n})|_{\Omega_{k}}\}_{n=1}^{\infty}$
is bounded in $H^{1}(\Omega_{k})\times H^{1}(\Omega_{k})$ and
hence precompact in
${}{L}^{2}(\Omega_{k})\times{}{L}^{2}(\Omega_{k})$. Therefore,
there is
$(\tilde{\xi},\tilde{\theta})\in{}{L}^{2}(\Omega_{k})\times{}{L}^{2}(\Omega_{k})$
such that, up to a subsequence,
\begin{equation}\label{65}
S(t_{n})(\xi_{0,n},\theta_{0,n})\longrightarrow(\tilde{\xi},\tilde{\theta})\quad\textrm{in}\quad{}{L}^{2}(\Omega_{k})\times{}{L}^{2}(\Omega_{k})\;.
\end{equation}
By (\ref{62}) and (\ref{65}), we find that
\begin{displaymath}
(\tilde{\xi},\tilde{\theta})=(\xi,\theta)|_{\Omega_{k}}\;,
\end{displaymath}
which means that for every $k\geq k_{1}$,
\begin{equation}\label{66}
S(t_{n})(\xi_{0,n},\theta_{0,n})|_{\Omega_{k}}\longrightarrow(\xi,\theta)|_{\Omega_{k}}\quad\textrm{in}\quad{}{L}^{2}(\Omega_{k})\times{}{L}^{2}(\Omega_{k})\;.
\end{equation}
In other words, for the given $\epsilon>0$, there is $N_{3}>0$
such that for all $k\geq k_{1}$ and $n\geq N_{3}$,
\begin{equation}\label{67}
\int_{\Omega_{k}}\left(|S(t_{n})\xi_{0,n}-\xi|^{2}+|S(t_{n})\theta_{0,n}-\theta|^{2}\right)dxdy\leq\frac{\epsilon}{5}\;.
\end{equation}
Since $\xi$ and $\theta$ are in ${}{L}^{2}(\Omega)$, there is
$k_{2}>0$ such that for all $k\geq k_{2}$,
\begin{equation}\label{68}
\int_{\Omega\setminus\Omega_{k}}\left(|\xi|^{2}+|\theta|^{2}\right)dxdy\leq\frac{\epsilon}{5}\;.
\end{equation}
Let $k_{0}=\max\{k_{1},k_{2}\}$ and
$N_{0}=\max\{N_{1},N_{2},N_{3}\}$, then for all $n\geq N$, we have
\begin{eqnarray}\label{69}
&\phantom{1}&\int_{\Omega}\left(|S(t_{n})\xi_{0,n}-\xi|^{2}+|S(t_{n})\theta_{0,n}-\theta|^{2}\right)dxdy\nonumber\\
&=&\int_{\Omega_{k_{0}}}\left(|S(t_{n})\xi_{0,n}-\xi|^{2}+|S(t_{n})\theta_{0,n}-\theta|^{2}\right)dxdy\nonumber\\
&+&\int_{\Omega\setminus\Omega_{k_{0}}}\left(|S(t_{n})\xi_{0,n}-\xi|^{2}+|S(t_{n})\theta_{0,n}
-\theta|^{2}\right)dxdy\nonumber\\
&\leq&\int_{\Omega_{k_{0}}}\left(|S(t_{n})\xi_{0,n}-\xi|^{2}+|S(t_{n})\theta_{0,n}-\theta|^{2}\right)dxdy\nonumber\\
&+&2\int_{\Omega\setminus\Omega_{k_{0}}}\left(|S(t_{n})\xi_{0,n}|^{2}+|S(t_{n})\theta_{0,n}|^{2}\right)dxdy\nonumber\\
&+&2\int_{\Omega\setminus\Omega_{k_{0}}}\left(|\xi|^{2}+|\theta|^{2}\right)dxdy \le \epsilon\;,
\end{eqnarray}
where the last inequality is obtained by (\ref{64}), (\ref{67})
and (\ref{68}). Notice that (\ref{69}) shows that
\begin{displaymath}
S(t_{n})(\xi_{0,n},\theta_{0,n})\longrightarrow(\xi,\theta)\quad\textrm{in}\quad{}{L}^{2}(\Omega)\times{}{L}^{2}(\Omega)\;,
\end{displaymath}
and hence $\{S(t)\}_{t\geq0}$ is asymptotically compact.
The proof is complete.
\end{proof}

We are, now, ready to prove the existence of a global attractor
for problem (\ref{7})-(\ref{8}).

\begin{theorem}\label{theorem:1}
Problem (\ref{7})-(\ref{8}) has a global attractor ${\mathcal{A}}$
in ${}{L}^{2}(\Omega)\times{}{L}^{2}(\Omega)$, which is compact,
invariant and attracts every bounded set with respect to the norm
of ${}{L}^{2}(\Omega)\times{}{L}^{2}(\Omega)$.
\end{theorem}

\begin{proof}
 By Lemma \ref{lemma:2},  the dynamical system
$\{S(t)\}_{t\geq0}$ has a bounded absorbing set in
${}{L}^{2}(\Omega)\times{}{L}^{2}(\Omega)$, and by Lemma
 \ref{lemma:6}, $\{S(t)\}_{t\geq0}$ is asymptotically compact.
Then the existence of a global attractor follows immediately from
the standard attractor theory (see e.g.,  \cite{bab1, bal1, hal1, sel1, tem1}).
\end{proof}

\section{Regularity of Global Attractors}
\setcounter{equation}{0}

In this section,   we investigate the regularity of the global
attractor obtained in Theorem  \ref{theorem:1}. We will show that
the global attractor ${\mathcal{A}}$ is actually contained in a bounded
subset of $H^{2}(\Omega)\times H^{2}(\Omega)$. We start with the
following lemma
\begin{lemma}\label{lemma:7}
Suppose that
$(\xi_{0},\theta_{0})\in{}{L}^{2}(\Omega)\times{}{L}^{2}(\Omega)$.
Then the solution $(\xi,\theta)$ of   problem
(\ref{7})-(\ref{8}) satisfies
\begin{displaymath}
\bigg\|\frac{d\xi}{dt}\bigg\|+\bigg\|\frac{d\theta}{dt}\bigg\|\leq
M,\quad\forall t\geq T\;,
\end{displaymath}
where $M$ depends only on the data $(\Omega,P_{r},R_{a})$, $T$
depends on the data $(\Omega,P_{r},R_{a})$ and $R$ when
$||\xi_{0}||\leq R$ and $||\theta_{0}||\leq R$
\end{lemma}

\begin{proof}
 By \eqref{one*}  and (\ref{7}) we find that
\begin{eqnarray}\label{70}
\bigg\|\frac{d\xi}{dt}\bigg\|&\leq&||\triangle\xi||+||J(\psi,\xi)||+C||\nabla\theta||+||f||\nonumber\\
&\leq&||\triangle\xi||+C||\xi||||\triangle\xi||+C_{1}||\nabla\theta||+C_{2}\leq
C||\triangle\xi||+C_{1},\quad\forall t\geq T\;,
\end{eqnarray}
where the last inequality is obtained by Lemma  \ref{lemma:3}.  By
(\ref{70}) and Lemma  \ref{lemma:3}  again we get, for $t\geq T$,
\begin{equation}\label{71}
\int_{t}^{t+1}\bigg\|\frac{d\xi}{dt}\bigg\|^{2}dt\leq
C\int_{t}^{t+1}||\triangle\xi||^{2}dt+C_{1}\leq C\;.
\end{equation}
Similarly, by (\ref{8}), we find that, for $t\geq T$,
\begin{equation}\label{72}
\bigg\|\frac{d\theta}{dt}\bigg\|\leq
C||\triangle\theta||+||J(\psi,\xi)||+||g||\leq
C||\triangle\theta||+C_{1}\;,
\end{equation}
which along with    Lemma  \ref{lemma:3} implies that,  for
$t\geq T$,
\begin{equation}\label{73}
\int_{t}^{t+1}\bigg\|\frac{d\theta}{dt}\bigg\|^{2}dt\leq
C\int_{t}^{t+1}||\triangle\theta||^{2}dt+C_{1}\leq C\;.
\end{equation}
Let $\tilde{\xi}=\frac{d\xi}{dt}$ and
$\tilde{\theta}=\frac{d\theta}{dt}$. Then it follows from
(\ref{71}) and (\ref{73}) that, for $t\geq T$,
\begin{equation}\label{74}
\int_{t}^{t+1}\left(||\tilde{\xi}(t)||^{2}+||\tilde{\theta}(t)||^{2}\right)dt\leq
C\;.
\end{equation}
We now differentiate (\ref{7}) and (\ref{8}) with respect to $t$
to obtain
\begin{equation}\label{75}
\frac{\partial\tilde{\xi}}{\partial
t}-\triangle\tilde{\xi}+J(\Psi_{t},\xi)+J(\Psi,\tilde{\xi})+\frac{R_{a}}{P_{r}}\frac{\partial\tilde{\theta}}{\partial
x}=0\;,
\end{equation}
and
\begin{equation}\label{76}
\frac{\partial\tilde{\theta}}{\partial
t}-\frac{1}{P_{r}}\triangle\tilde{\theta}+J(\Psi_{t},\theta)+J(\Psi,\tilde{\theta})=0\;.
\end{equation}
Taking the inner product of (\ref{75}) with $\tilde{\xi}$ in
${}{L}^{2}(\Omega)$, we find that
\begin{equation}\label{77}
\frac{1}{2}\frac{d}{dt}||\tilde{\xi}||^{2}+||\nabla\tilde{\xi}||^{2}+\Big(J(\Psi_{t},\xi),\tilde{\xi}\Big)+\frac{R_{a}}{P_{r}}\left(\frac{\partial\tilde{\theta}}{\partial
x},\tilde{\xi}\right)=0\;.
\end{equation}
By \eqref{one*}  we have
\begin{equation}\label{78}
\left|\Big(J(\Psi_{t},\xi),\tilde{\xi}\Big)\right|\leq||J(\Psi_{t},\xi)||||\tilde{\xi}||\leq
C||\tilde{\xi}||^{2}||\triangle\xi||\leq||\tilde{\xi}||^{4}+C||\triangle\xi||^{2}\;.
\end{equation}
We also have the following inequality
\begin{equation}\label{79}
\left|\frac{R_{a}}{P_{r}}\left(\frac{\partial\tilde{\theta}}{\partial
x},\tilde{\xi}\right)\right|\leq
C||\nabla\tilde{\theta}||||\tilde{\xi}||\leq\frac{1}{2P_{r}}||\nabla\tilde{\theta}||^{2}+C||\tilde{\xi}||^{2}\leq\frac{1}{2P_{r}}||\nabla\tilde{\theta}||^{2}+||\tilde{\xi}||^{4}+C_{1}\;.
\end{equation}
It follows from (\ref{77})-(\ref{79}) that
\begin{equation}\label{80}
\frac{d}{dt}||\tilde{\xi}||^{2}+||\nabla\tilde{\xi}||^{2}\leq
C||\tilde{\xi}||^{4}+C_{1}||\triangle\xi||^{2}+\frac{1}{2P_{r}}||\nabla\tilde{\theta}||^{2}+C_{2}\;.
\end{equation}
Now, by taking the inner product of (\ref{76}) with
$\tilde{\theta}$ in ${}{L}^{2}(\Omega)$ we get
\begin{equation}\label{81}
\frac{1}{2}\frac{d}{dt}||\tilde{\theta}||^{2}+\frac{1}{P_{r}}||\nabla\tilde{\theta}||^{2}=-\left(J(\Psi_{t},\theta),\tilde{\theta}\right)\;.
\end{equation}
By an argument similar to (\ref{78}), the right-hand side of
(\ref{81}) is bounded by
\begin{equation}\label{82}
\left|\left(J(\Psi_{t},\theta),\tilde{\theta}\right)\right|\leq||\tilde{\xi}||^{4}+||\tilde{\theta}||^{4}+C||\triangle\theta||^{2}\;.
\end{equation}
By (\ref{81}) and (\ref{82}) we find that
\begin{equation}\label{83}
\frac{d}{dt}||\tilde{\theta}||^{2}+\frac{1}{P_{r}}||\nabla\tilde{\theta}||^{2}\leq
2||\tilde{\xi}||^{4}+2||\tilde{\theta}||^{4}+C||\triangle\theta||^{2}\;.
\end{equation}
By (\ref{80}) and (\ref{83}) it follows that
$$
\frac{d}{dt}\left(||\tilde{\xi}||^{2}+||\tilde{\theta}||^{2}\right)+||\nabla\tilde{\xi}||^{2}+\frac{1}{2P_{r}}||\nabla\tilde{\theta}||^{2}
$$
\begin{equation}\label{84}
\leq
C\left(||\tilde{\xi}||^{4}+||\tilde{\theta}||^{4}\right)+C_{1}\left(1+||\triangle\xi||^{2}+||\triangle\theta||^{2}\right)\;,
\end{equation}
which implies that
\begin{eqnarray}\label{85}
\frac{d}{dt}\left(||\tilde{\xi}||^{2}+||\tilde{\theta}||^{2}\right)&\leq&C\left(||\tilde{\xi}||^{2}
+||\tilde{\theta}||^{2}\right)\left(||\tilde{\xi}||^{2}+||\tilde{\theta}||^{2}\right)\nonumber\\
&+&C_{1}\left(1+||\triangle\xi||^{2}+||\triangle\theta||^{2}\right)\;.
\end{eqnarray}
By Lemma  \ref{lemma:3},  (\ref{74}) and the uniform Gronwall
inequality we finally obtain that
\begin{equation}\label{86}
||\tilde{\xi}(t)||^{2}+||\tilde{\theta}(t)||^{2}\leq
C,\quad\forall t\geq T+1\;,
\end{equation}
which concludes the proof.
\end{proof}

\begin{lemma}\label{lemma:8}
Suppose that
$(\xi_{0},\theta_{0})\in{}{L}^{2}(\Omega)\times{}{L}^{2}(\Omega)$.
Then the solution $(\xi,\theta)$ of  problem
(\ref{7})-(\ref{8}) satisfies
\begin{displaymath}
||\xi(t)||_{H^{2}}+||\theta(t)||_{H^{2}}\leq M,\quad\forall t\geq
T\;,
\end{displaymath}
where $M$ depends only on the data $(\Omega,P_{r},R_{a})$, $T$
depends on the data $(\Omega,P_{r},R_{a})$ and $R$ when
$||\xi_{0}||\leq R$ and $||\theta_{0}||\leq R$
\end{lemma}

\begin{proof}
 By (\ref{7}) we have the following inequality
\begin{eqnarray}\label{87}
||\triangle\xi||&\leq&\left\|\frac{\partial\xi}{\partial
t}\right\|+||J(\psi,\xi)||+C||\nabla\theta||+||f||\leq
C+C_{1}||\Psi||_{H^{3}}||\nabla\xi||\nonumber\\
&\leq&C+C_{1}||\nabla\xi||^{2}\leq C\;,
\end{eqnarray}
where we have used \eqref{two*}  and Lemmas  \ref{lemma:3}  and
 \ref{lemma:7}. Similarly, by (\ref{7}) we see that
\begin{equation}\label{88}
||\triangle\theta||\leq\left\|\frac{\partial\theta}{\partial
t}\right\|+||J(\psi,\xi)||+||g||\leq C\;.
\end{equation}
From (\ref{87}) and (\ref{88}), Lemma  \ref{lemma:8} follows.
\end{proof}

We are  now  in position to prove the regularity of the global
attractor in $H^{2}(\Omega)\times H^{2}(\Omega)$.

\begin{theorem}\label{theorem:2}
The global attractor ${\mathcal{A}}$ obtained in Theorem  \ref{theorem:1}
is bounded in $H^{2}(\Omega)\times H^{2}(\Omega)$, i.e., there is
a positive constant $M$ such that
\begin{displaymath}
||(\xi,\theta)||_{H^{2}(\Omega)\times H^{2}(\Omega)}\leq
M,\quad\forall (\xi,\theta)\in{}{A}\;.
\end{displaymath}
\end{theorem}

\begin{proof}
 Since ${}{A}$ is bounded in
${}{L}^{2}(\Omega)\times{}{L}^{2}(\Omega)$, by Lemma
 \ref{lemma:8},  there is a bounded set $E\subseteq
H^{2}(\Omega)\times H^{2}(\Omega)$ and $T>0$ such that
\begin{displaymath}
S(t){}{A}\subseteq E,\quad\forall t\geq T\;.
\end{displaymath}
But $S(t){}{A}\subseteq{}{A}$ and hence ${}{A}\subseteq E$. The
proof is complete.
\end{proof}


\begin{thebibliography}{100}


\bibitem{ant1}
F. Antoci and M. Prizzi, Reaction-Diffusion equations on unbounded
thin domains,
{\em Topological Methods in Nonlinear  Analysis},
{\bf 18} (2001), 283-302.


\bibitem{ant2}
F. Antoci and M. Prizzi, Attractors and global averaging of non-autonomous
Reaction-Diffusion equations in $\mathbb{R}^n$.
{\em Topological Methods in Nonlinear  Analysis},
{\bf 20} (2002), 229-259.



\bibitem{arr1}
J. M. Arrieta, J. W. Cholewa, T. Dlotko,  and A. Rodriguez-Bernal,
Asymptotic behavior and attractors for Reaction Diffusion equations
in unbounded  domains,
{\em Nonlinear Analysis}, {\bf 56} (2004), 515-554.




\bibitem{bab1}
A.V. Babin and  M.I.  Vishik,     Attractors of
Evolution Equations,  North-Holland, Amsterdam,
1992.


\bibitem{bal1}
J.M.   Ball,  Continuity properties and global
attractors of generalized semiflows and the Navier-Stokes
equations, {\em  J. Nonl. Sci.},  {\bf 7} (1997),
475-502.


\bibitem{bal2}
J.M.   Ball,
Global attractors for damped semilinear wave equations,
  {\em  Discrete Contin.  Dyn.  Syst.}   {\bf 10}   (2004),
31-52.




\bibitem{che1}
S. Chen, Symmetry analysis of convection on patterns,
{\em Comm. Theor. Phys.}, {\bf 1} (1982), 413-426.

\bibitem{fei1}
M. J. Feigenbaum, The onset spectrum of turbulence,
{\em Phys. Lett. A}, {\bf 74} (1979), 375-378.

\bibitem{guo1}
B. Guo, Spectral method for solving the two-dimensional New-Boussinesq equations,
{\em Acta. Math. Appl.}, {\bf 5} (1989), 208-218.

\bibitem{guo2}
B. Guo,  Nonlinear Galerkin methods for solving the
   two-dimensional New-Boussinesq equations,
{\em Advance in Math.}, {\bf 22} (1993), 179-181.

\bibitem{guo3}
B. Guo and B. Wang, Gevrey class regularity and approximate inertial manifolds
for the New-Boussinesq equations, {\em Chin. Ann. of Math.}, {\bf 19B} (1998), 179-188.

\bibitem{guo4}
B. Guo and B. Wang, Approximate inertial manifolds
for the  two-dimensional New-Boussinesq equations, {\em J. Partial Diff. Equ.},
{\bf  9} (1996),  237-250.





 \bibitem{hal1}
  J.K.  Hale,   Asymptotic Behavior of
Dissipative Systems,
American Mathematical Society,
  Providence, RI, 1988.



\bibitem{ju1}
N. Ju,  The $H^1$-compact global attractor for the solutions
to the Navier-Stokes equations in two-dimensional unbounded domains,
  {\em  Nonlinearity},  {\bf 13} (2000),
1227-1238.

 \bibitem{mor1}
 F. Morillas and J. Valero,
 Attractors for Reaction-Diffusion equations in $\mathbb{R}^n$ with
 continuous  nonlinearity, {\em Asymptotic Analysis}, {\bf 44} (2005),
 111-130.

\bibitem{moi1}
I. Moise, R. Rosa and X. Wang,
Attractors for non-compact semigroups via energy equations,
  {\em  Nonlinearity},  {\bf 11} (1998),  1369-1393.



\bibitem{pri1}
M. Prizzi, Averaging, Conley index continuation and recurrent
dynamics in almost-periodic equations,
{\em J. Differential Equations}, {\bf 210} (2005), 429-451.


    \bibitem{rob1}
 J.C. Robinson, Infinite-Dimensional Dynamical Systems,
 Cambridge University Press, Cambridge, UK, 2001.


\bibitem{rod1}
A. Rodrigue-Bernal and B. Wang,
Attractors for partly dissipative Reaction Diffusion systems
in $\mathbb{R}^n$,
{\em J. Math. Anal. Appl.}, {\bf 252} (2000),  790-803.

 \bibitem{ros1}
R. Rosa,  The global attractor for the 2D Navier-Stokes flow on some
unbounded  domains,
  {\em  Nonlinear Anal.},  {\bf 32} (1998),   71-85.


    \bibitem{sel1}
 R. Sell and  Y. You,
 Dynamics of Evolutionary Equations,
 Springer-Verlag,
New York, 2002.



 \bibitem{sun1}
 C. Sun and C. Zhong, Attractors for the semilinear Reaction-Diffusion
 equation with distribution derivatives in unbounded  domains,
 {\em Nonlinear Analysis}, {\bf 63} (2005), 49-65.


\bibitem{tem1}
 R.  Temam,    Infinite-Dimensional Dynamical
Systems in Mechanics and Physics,
  Springer-Verlag,
  New York, 1997.




\bibitem{wan1}
B. Wang,  Attractors for reaction-diffusion equations in unbounded domains,
{\em Physica D}, {\bf 128}  (1999), 41-52.






\end{thebibliography}
\end{document}